\documentclass[fleqn,usenatbib]{mnras}

\usepackage{newtxtext,newtxmath}

\usepackage[T1]{fontenc}

\DeclareRobustCommand{\VAN}[3]{#2}
\let\VANthebibliography\thebibliography
\def\thebibliography{\DeclareRobustCommand{\VAN}[3]{##3}\VANthebibliography}

\usepackage[latin9]{inputenc}
\usepackage{verbatim}
\usepackage{graphicx}	
\usepackage{amsmath}	
\usepackage{amsbsy}
\usepackage{amstext}
\usepackage{babel}

\title[\texttt{iSyMBA}]{\texttt{iSyMBA}: A Symplectic Massive Bodies Integrator with Planets Interpolation}

\author[F. Roig et al.]{
Fernando Roig,$^{1}$\thanks{E-mail: froig@on.br}
David Nesvorn\'{y},$^{2}$
Rogerio Deienno$^{2}$
and Matias J. Garcia$^{1}$
\\
$^{1}$Observat\'{o}rio Nacional, Rua Gal. Jose Cristino 77, Rio de Janeiro, RJ 20921-400, Brazil\\
$^{2}$Department of Space Studies, Southwest Research Institute, 1050 Walnut St., Suite 300, Boulder, CO 80302, USA
}

\date{Accepted XXX. Received YYY; in original form ZZZ}

\pubyear{2021}

\begin{document}
\label{firstpage}
\pagerange{\pageref{firstpage}--\pageref{lastpage}}
\maketitle

\begin{abstract}
A planetary instability occurring at time $<100$ My after formation of
the giant planets in our solar system can be responsible for some characteristics of the inner solar system.
However, the actual influence of the instability on
the terrestrial planet formation is not well understood. The simulations of terrestrial planet formation are very CPU-expensive, and this limits
the exploration of different instability scenarios. To include
the effects of the giant planets instability in the
simulations of terrestrial planets formation in a feasible way, we
approach the problem in two steps. First, we model and
record an evolution of the giant planets that replicates
the present outer solar system in the end. Then, we use that orbital record, properly interpolated, as the input for a second step to simulate its effects on the terrestrial planet formation. For
this second step, we developed \texttt{iSyMBA}, a symplectic massive
bodies algorithm, where ``i'' stands for interpolation. \texttt{iSyMBA}
is a very useful code to accurately evaluate the effects of planetary
instabilities on minor
body reservoirs, while accounting for close encounters among massive
objects. We provide a detailed description of how \texttt{iSyMBA}
was developed and implemented to study terrestrial planet formation.
Adapting \texttt{iSyMBA} for other problems that demand interpolation
from previous simulations can be done following the method described here.
\end{abstract}



\section{Introduction}

Current theories of the early evolution of the solar system invoke
a temporary instability of the giant planets, that would have happened
sometime after the dissipation of the gas in the protoplanetary nebula
\citep[see][for a review]{Nesvorny2018}. This instability led to
mutual scattering of the giant planets while they were radially
migrating due to the interaction with an outer massive disk of remnant
planetesimals. The instability is required
to explain many dynamical features that are currently observed in
the different populations of solar system bodies. These include the
angular momentum deficit of the giant planets \citep[e.g.][]{Nesvorny2012,Deienno2017},
the inclinations of asteroids in the main belt \citep[e.g.][]{Roig2015,Deienno2018},
the existence of Jupiter trojans \citep[e.g.][]{Nesvorny2013}, the
orbital architecture of the Kuiper belt \citep[e.g.][]{Nesvorny2015,Gomes2018},
the excited orbit of Mercury \citep[e.g.][]{Roig2016}, the dynamical
characteristics of the satellites of the jovian planets \citep[e.g.][]{Deienno2014,Nesvorny2014,Nesvorny2014b},
among others.

The first instability models \citep{Tsiganis2005} assumed that the
instability may have happened around 600 My after the dissipation
of the gas in the protoplanetary nebula, helping to trigger the Lunar
Late Heavy Bombardment \citep[LHB;][]{Gomes2005,Bottke2012}. Recent
models, however, propose that the instability may have happened as
early as $\sim10$ My after the dissipation of the gas \citep{Nesvorny2018}.
This means that the instability might have played a relevant role
in the accretion of the terrestrial planets \citep[e.g.][]{Clement2018,Clement2019,Nesvorny2021}.

Using fully self consistent models to study the effects of the instability
in the early evolution of the solar system may be an unfeasible task.
In general, such models need to consider at least three ingredients:
the giant planets, the disk of massive planetesimals that drives the
migration of the giant planets, and the population of bodies affected
by the instability. We refer to the latter as the \textsl{target population},
and it may be represented either by test particles or by mutually interacting massive bodies. This requires to explore a large number of model
parameters and over a wide range of values, implying the need for hundreds
or thousands of computationally expensive numerical simulations that,
in most cases, lead to meaningless results.

An alternative to overcome these limitations is to use simplified,
yet realistic, models where the amount of free parameters is smaller,
and the user have more control over the range of possible evolutions
of the system. This involves, for example, to drop off the disk of
planetesimals and consider a prescribed evolution of the giant planets.
This prescribed evolution might be as simple as an \textit{ad hoc} evolution,
mimicked by using artificial forces, or as complex as a realistic
evolution obtained from other previous simulations \citep[e.g.][]{Nesvorny2017AJ,Deienno2018}. 

Here, we develop a numerical integration method that exploits the
latter approach. In this method, the evolution of the giant planets
is stored in a file at regular intervals, and their positions and
velocities for a given time are obtained by interpolation. Our aim
is to construct a symplectic $N$ body integrator for the target population,
that has the giant planets interpolation scheme embedded in it. 

Symplectic integrators are a particular class of numerical integrators,
specifically designed to solve Hamiltonian problems \citep{Yoshida1993}.
Their main property is the conservation of a quantity $\overline{H}$, referred to as the surrogate Hamiltonian,
that is very close to the original Hamiltonian $H$ of the problem. Our choice
for a symplectic algorithm has two motivations: (i) these algorithms
have proven to be fast and reliable, allowing for long term simulations
of planetary $N$-body systems with less computational cost than traditional
integration algorithms \citep[e.g.][]{Wisdom1991}, and (ii) we intend
to use the set of subroutines and packages of the well-tested and publicly
available symplectic integrators \texttt{Swift} \citep{Levison1993}
and \texttt{SyMBA} \citep{Duncan1998}, with the necessary modifications, as the basis for our algorithm.

When the target population consists of mass-less particles, the construction of a symplectic integrator with embedded interpolation is relatively straightforward \citep[e.g.][]{Beauge2002,Roig2015}.
On the other hand, when the target populations consist of massive bodies that interact with each other and may develop close encounters,
including the giant planets interpolation in a symplectic way requires some specific considerations that turn the development of the algorithm more difficult. One particular problem is related to the fact that the system of giant planets has a centre-of-mass that differs from that of the target population. This requires a specific splitting of the Hamiltonian that is not accounted for in full N-body symplectic integrators.

Here, we focus on the case when the target population is in the inner solar system, and we describe the construction of the algorithm. For its recent applications, we refer the reader to \citet{Nesvorny2021} and \citet{DeSouza2021}.
The paper is organised as follows. In Sect. \ref{sect2}, we revise the basic concepts, provide the detailed description of the new algorithm, and perform some validation tests. In Sect. \ref{setc3}, we briefly discuss the possible parallelization strategies for the new code. The last section is devoted to the conclusions.

\section{Symplectic integration with interpolation}\label{sect2}

\subsection{\texttt{SyMBA}}

\begin{figure*}
\centering
\includegraphics[width=\textwidth]{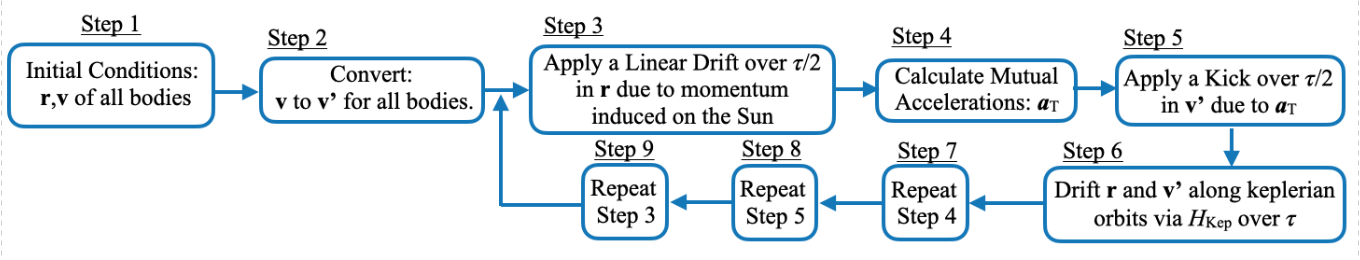}
\caption{The detailed sequence of operations needed for a single integration time step, from $t$ to $t+\tau$ in the \texttt{SyMBA} code. Primed variables
are referred to the centre-of-mass of the system composed by the giant planets, the terrestrial proto-planets, and the terrestrial planetesimals.
Heliocentric variables are not primed.}
\label{fig1}
\end{figure*}

Before proceeding with the description of \texttt{iSyMBA}, we will
briefly review the basics of \texttt{SyMBA} \citep[see][for more details]{Duncan1998}. \texttt{SyMBA}
stands for Symplectic Massive Bodies Algorithm, and it is a second
order symplectic integrator for the planetary $N$ body problem, that
allows for close encounters between massive bodies. 

Let us assume a set of $N$ bodies of masses $m_{i}$ orbiting around
the Sun (or any central star) of mass $M\gg m_{i}$, and introduce the Poincar\'{e} canonical coordinates
$\mathbf{r}_{i},\mathbf{p}_{i}^{\prime}$ such that $\mathbf{r}_{i}$
are heliocentric positions and $\mathbf{p}_{i}^{\prime}=m_{i}\mathbf{v}_{i}^{\prime}$
are barycentric momenta (barycentric velocities). Throughout this work, primed variables are referred to the centre-of-mass
of the system, and not primed variables are referred to the Sun. The Hamiltonian of the system is then composed of three parts:
\begin{equation}
H(\mathbf{r},\mathbf{p}^{\prime})=H_{\mathrm{Kep}}+H_{\mathrm{Int}}+H_{\mathrm{Sun}}\label{eq:ap0}
\end{equation}
where
\begin{equation}
H_{\mathrm{Kep}}(\mathbf{r},\mathbf{p}^{\prime})=\sum_{i=1}^{N}\left(\frac{\left|\mathbf{p}_{i}^{\prime}\right|^{2}}{2m_{i}}-\frac{GMm_{i}}{\left|\mathbf{r}_{i}\right|}\right)\label{eq:ap3-1}
\end{equation}
represents the two body motion of the bodies around the Sun,
\begin{equation}
H_{\mathrm{Int}}(\mathbf{r})=-\sum_{i=1}^{N-1}\sum_{k=i+1}^{N}\frac{Gm_{i}m_{k}}{\left|\mathbf{r}_{i}-\mathbf{r}_{k}\right|}\label{eq:ap4-1}
\end{equation}
is the gravitational interaction potential between the bodies, and
\begin{equation}
H_{\mathrm{Sun}}(\mathbf{p}^{\prime})=\frac{1}{2M}\left|\sum_{i=1}^{N}\mathbf{p}_{i}^{\prime}\right|^{2}\label{eq:ap6-1}
\end{equation}
is the barycentric kinetic energy of the Sun. 

Whenever $H_{\mathrm{Int}},H_{\mathrm{Sun}}\ll H_{\mathrm{Kep}}$,
a second order symplectic integration over a time step $\tau$ is
obtained through the following sequence of steps:
\begin{description}
\item Step 1. Evolve the system only through $H_{\mathrm{Sun}}$, over $\tau/2$;
this applies a linear drift (LD) to the positions $\mathbf{r}$.
\item Step 2. Evolve the system only through $H_{\mathrm{Int}}$, over $\tau/2$;
this applies an impulse or kick (K) to the momenta $\mathbf{p}^{\prime}$.
\item Step 3. Evolve the system only through $H_{\mathrm{Kep}}$, over $\tau$;
this applies a drift (D) to each body along a Keplerian orbit.
\item Step 4. Evolve the system only through $H_{\mathrm{Int}}$, over $\tau/2$;
this applies again a kick (K) to the momenta $\mathbf{p}^{\prime}$.
\item Step 5. Evolve the system only through $H_{\mathrm{Sun}}$, over $\tau/2$; this applies again a linear drift (LD) to the positions $\mathbf{r}$.
\end{description}
The integration over a time step $\tau$ is then schematised
by a sequence LD - K - D - K - LD. The detailed flowchart of the algorithm
is shown in Fig. \ref{fig1}.

However, when a close encounter between two bodies arises, the condition
$H_{\mathrm{Int}}\ll H_{\mathrm{Kep}}$ is no longer satisfied
and the above sequence is no longer valid. In a traditional numerical
integrator, the increase of the $H_{\mathrm{Int}}$ term during the
close encounter is usually compensated by a decrease of the time step
$\tau$, such as to keep the impulse $-\tau\,\partial H_{\mathrm{Int}}/\partial\mathbf{r}$
limited. However, in a symplectic integrator the time step must be
kept fixed over the whole integration. This is a well known limitation
of symplectic algorithms, since the surrogate Hamiltonian $\overline{H}$,
that is preserved by the algorithm, is such that $\overline{H}=H+\mathcal{O}(\tau^{n})$,
being $n$ the order of the integrator. Thus, any change in the time
step would break the conservation of $\overline{H}$.

The solution proposed by \citet{Duncan1998} and \citet{Chambers1999}
to circumvent this limitation consists into split the term $H_{\mathrm{Int}}$
into pieces, weighted by a function that depends on the distance between
the different pairs of bodies. Consider, for example, a two-terms splitting, $H_{\mathrm{Int}}=H_{\mathrm{Int}}^{(0)}+H_{\mathrm{Int}}^{(1)}$,
where:
\begin{align}
H_{\mathrm{Int}}^{(0)} & =-\sum_{i=1}^{N-1}\sum_{k=i+1}^{N}\frac{Gm_{i}m_{k}}{r_{ik}}\left(1-w_{1,ik}\right)\nonumber \\
H_{\mathrm{Int}}^{(1)} & =-\sum_{i=1}^{N-1}\sum_{k=i+1}^{N}\frac{Gm_{i}m_{k}}{r_{ik}}w_{1,ik}\label{eq:ap18}
\end{align}
Here, $r_{ik}=\left|\mathbf{r}_{i}-\mathbf{r}_{k}\right|$, and $w_{l,ik}$
is a sigmoid-like weight function:
\begin{equation}
w_{l,ik}=\left\{ \begin{array}{ll}
1 & ,\qquad r_{ik}\geq R_{l,ik}\\
\phi\left(\frac{r_{ik}-R_{l+1,ik}}{R_{l,ik}-R_{l+1,ik}}\right) & ,\qquad R_{l+1,ik}\leq r_{ik}<R_{l,ik}\\
0 & ,\qquad r_{ik}<R_{l+1,ik}
\end{array}\right.\label{eq:ap19-2}
\end{equation}
where $\phi$ is a suitable odd degree polynomial, and $R_{l,ik},R_{l+1,ik}$
are defined in terms of the mutual Hill radii. Then, the sequence of steps
over a time step $\tau$ becomes: 
\begin{description}
\item Step 1. Evolve the system only through $H_{\mathrm{Sun}}$, over $\tau/2$
\item Step 2. Evolve the system only through $H_{\mathrm{Int}}^{(1)}$, over $\tau/2$
\item Step 3. Repeat from 1 to $q$ ($q>1$, integer)
\item \quad{}Step 3.1. Evolve the system only through $H_{\mathrm{Int}}^{(0)}$, over
$\tau/2q$
\item \quad{}Step 3.2. Evolve the system only through $H_{\mathrm{Kep}}$, over $\tau/q$
\item \quad{}Step 3.3. Evolve the system only through $H_{\mathrm{Int}}^{(0)}$, over
$\tau/2q$
\item Step 4. Evolve the system only through $H_{\mathrm{Int}}^{(1)}$, over $\tau/2$
\item Step 5. Evolve the system only through $H_{\mathrm{Sun}}$, over $\tau/2$
\end{description}
When $r_{ik}\geq R_{l,ik}$ (no close encounters), $H_{\mathrm{Int}}^{(0)}=0$
and step 3 reduces to a single evolution of $H_{\mathrm{Kep}}$ over
$\tau$. On the other hand, when $r_{ik}<R_{l+1,ik}$ (close encounter),
$H_{\mathrm{Int}}^{(1)}=0$, and step 3 effectively performs the symplectic integration
using a smaller time step $\tau/q$. The above sequence can be represented
as:
\[
\begin{array}{ccc}
\underbrace{\text{LD - K}^{(1)}}\text{ -} 
\left(\underbrace{\mathrm{K}^{(0)}\text{-D-K}^{(0)}}\right)^{q} 
\text{- } \underbrace{\mathrm{K}^{(1)}\text{ - LD}}\\
{\scriptstyle \text{step }\tau} \qquad\qquad 
{\scriptstyle \text{step }\tau/q} \qquad\qquad
{\scriptstyle \text{step }\tau}
\end{array}
\]

This strategy can be recursively extended to an arbitrary sequence
of weight functions for radii $R_{1,ik}>R_{2,ik}>\ldots>R_{n,ik}>R_{n+1,ik}$,
associating smaller and smaller time steps to each weight function.
In \texttt{SyMBA}, $R_{n+1,ik}$ is usually set to be of the order
of the planetary radii, and $R_{1,ik}$ is of the order of a few Hill
radii.

\subsection{\texttt{iSyMBA}}

In order to apply the above concepts to develop \texttt{iSyMBA}, we
consider three different categories of bodies:
\begin{enumerate}
\item Giant planets: represented by $J$ bodies of masses $\mu_{j}$, $j=1,\ldots,J$.
The positions $\boldsymbol{\rho}_{j}$ and velocities $\boldsymbol{\upsilon}_{j}$
of these bodies are directly read from a file, where they are stored
at regular time intervals, and are interpolated down to the desired
time step.\footnote{The storing cadence needs to be dense enough for the interpolation to work properly. A cadence of 1 year proved to be good for interpolation to time steps of a few days. In principle, interpolation could be avoided by using a cadence equal to the time step, but this unnecessarily increases the file size and also makes the algorithm too slow due to the amount of I/O operations.}
The giant planets perturb the other two bodies categories
in the simulation, but they do not suffer any perturbation, neither
among them, nor from other bodies.
\item Terrestrial protoplanets: represented by $T$ bodies of masses $m_{i}\geq m_{\mathrm{tiny}}$,
$i=1,\ldots,T$. The positions $\mathbf{r}_{i}$ and velocities $\mathbf{v}_{i}$
of these bodies are advanced through a second order symplectic integrator.
They feel their mutual gravitational perturbations, as well as those
from the giant planets and from the terrestrial planetesimals. Terrestrial
protoplanets may also feel relativistic perturbations, if required.
\item Terrestrial planetesimals: represented by $N-T$ bodies of masses
$m_{i}<m_{\mathrm{tiny}}$, $i=T+1,\ldots,N$. The positions $\mathbf{r}_{i}$
and velocities $\mathbf{v}_{i}$ of these bodies are also advanced
through a second order symplectic integrator. They do not feel their
mutual gravitational perturbations, but they are perturbed by both
the giant planets and the terrestrial protoplanets. They also perturb the terrestrial protoplanets.
\end{enumerate}
As in the original \texttt{SyMBA} code, $m_{\mathrm{tiny}}$ is a
constant mass threshold, that we set, for example, to $2\,M_{\mathrm{Moon}}$ in \citet{Nesvorny2021}. We
refer to the set of terrestrial protoplanets and planetesimals simply
as the terrestrial bodies. 

\begin{figure*}
\centering
\includegraphics[width=\textwidth]{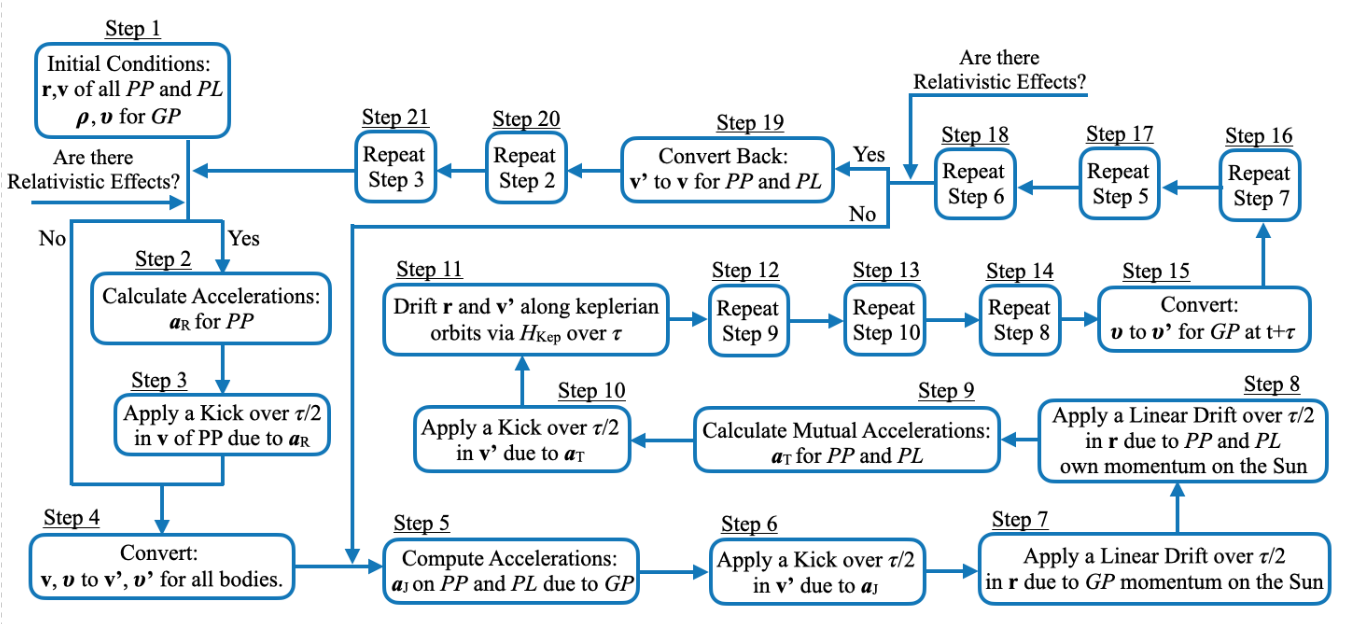}
\caption{The detailed sequence of operations needed for a single time step
integration, from $t$ to $t+\tau$ in the \texttt{iSyMBA} code. Primed variables
are referred to the centre-of-mass of the system composed by the giant planets (GP), the terrestrial proto-planets (PP) and planetesimals (PL).
Heliocentric variables are not primed.}
\label{fig2}
\end{figure*}

In the following, we describe in detail
the different parts of the code. The flowchart of the algorithm is presented in Fig. \ref{fig2}.

\subsubsection{Giant planets interpolation}\label{interp}

The positions and velocities of the giant planets are obtained from
a previous, and independent, simulation of the migration of these
planets by interaction with a transneptunian disk of planetesimals
\citep[e.g.][]{Nesvorny2012,Deienno2018}. The output of this simulation, i.e. heliocentric positions and velocities (or heliocentric orbital elements) of the giant planets, as
well as their masses and the Sun mass, are stored in a file at 1-yr intervals. The masses are not necessarily constant during such evolution, since the planets and the Sun may accrete planetesimals while migrating and grow in mass.

Let $\boldsymbol{\rho}_{b},\boldsymbol{\upsilon}_{b},\mu_{b}$ be
the position, velocity, and mass of a given giant planet at time $t_{b}$,
and $\boldsymbol{\rho}_{e},\boldsymbol{\upsilon}_{e},\mu_{e}$ the
corresponding values at a posterior time $t_{e}$. Let also $M_{b}$
and $M_{e}$ the corresponding masses of the Sun. Assume that we want
to interpolate this trajectory over a time step $\tau=(t_{e}-t_{b})/n$.
The interpolation method consists of the following steps
\begin{description}
\item Step 1. Advance $\boldsymbol{\rho}_{b},\boldsymbol{\upsilon}_{b}$ from $t_{b}$
to $t_{e}$ along a Keplerian orbit, considering the central mass
$M_{b}+\mu_{b}$, to obtain the sequence of values $\boldsymbol{\rho}_{b}^{(i)},\boldsymbol{\upsilon}_{b}^{(i)}$,
for times $t_{i}=t_{b}+i\tau$, $i=0,\ldots,n$. 
\item Step 2. Recede $\boldsymbol{\rho}_{e},\boldsymbol{\upsilon}_{e}$ from $t_{e}$
to $t_{b}$ along a Keplerian orbit, considering the central mass
$M_{e}+\mu_{e}$, to obtain the sequence of values $\boldsymbol{\rho}_{e}^{(j)},\boldsymbol{\upsilon}_{e}^{(j)}$,
for times $t_{j}=t_{e}-j\tau$, $j=0,\ldots,n$. 
\item Step 3. Compute the interpolated values at time $t_{b}\leq t_{k}\leq t_{e}$ through a weighted average:
\begin{align}
\boldsymbol{\rho}_{k} & 
=(1-w_{k})\boldsymbol{\rho}_{b}^{(k)}+w_{k}\boldsymbol{\rho}_{e}^{(n-k)}\nonumber \\
\boldsymbol{\upsilon}_{k} & =(1-w_{k})\boldsymbol{\upsilon}_{b}^{(k)}+w_{k}\boldsymbol{\upsilon}_{e}^{(n-k)}\nonumber \\
\boldsymbol{\mu}_{k} & 
=(1-w_{k})\boldsymbol{\mu}_{b}+w_{k}\boldsymbol{\mu}_{e}\nonumber \\
M_{k} & 
=(1-w_{k})M_{b}+w_{k}M_{e}
\label{eq:ap1}
\end{align}
where $w_{k}=k/n$, $k=0,\ldots,n$. 
\end{description}

We recall that these interpolated
coordinates are heliocentric. A similar interpolation strategy was already implemented in the \texttt{Swift\_RMVS3} integrator to deal with encounters of test particles to massive bodies \citep{Levison1993}, and the corresponding subroutine has been adapted with the necessary modifications to account for mass changes.

The interpolation of a giant planet that disappears from the system, either by escaping or by merging to another giant, can be performed only until the last registry of that planet stored in the file. In such case, since the orbits are stored at 1-yr intervals, we loose only a few months of evolution of the giant that disappears. This is not expected to have any significant influence on the evolution of the target population over million years. Moreover, in the case of a merging between two giant planets, one giant disappears but another has its mass increased. This latter giant will be properly interpolated since the interpolation scheme takes into account any variation of the planet's mass.

It is worth noting that the symplectic algorithm described in the next sections is independent of the particular interpolation scheme. Other interpolation routines, different that the one presented here, can be applied with the same result.

\subsubsection{Terrestrial bodies integration }

The key contribution to the development of the algorithm consists in the second order symplectic integrator to advance the orbits of the
terrestrial bodies over a time step $\tau$. This is constituted by a specific sequence of Lie series, applied to a non autonomous Hamiltonian of the form:
\begin{equation}
H(\mathbf{r},\mathbf{p}^{\prime},t)=H_{\mathrm{Kep}}+H_{\mathrm{Pert}}+H_{\mathrm{Jov}}+H_{\mathrm{Sun}}\label{eq:ap2}
\end{equation}
where, again, $\mathbf{r}_{i},\mathbf{p}_{i}^{\prime}$ are the Poincar\'{e} canonical
coordinates. The Hamiltonian is time dependent though the
heliocentric positions $\boldsymbol{\rho}_{j}$ and the barycentric
momenta (velocities) $\boldsymbol{\pi}_{j}^{\prime}=\mu_{j}\boldsymbol{\upsilon}_{j}^{\prime}$
of the giant planets, which are computed from the interpolated heliocentric
values (Sect. \ref{interp}). It is worth noting that the barycentric momenta $\mathbf{p}^{\prime},\boldsymbol{\pi}^{\prime}$
are referred to the centre-of-mass of the whole system, i.e. considering
the terrestrial bodies and the giant planets altogether.

The different terms of the Hamiltonian are:
\begin{equation}
H_{\mathrm{Kep}}(\mathbf{r},\mathbf{p}^{\prime})=\sum_{i=1}^{N}\left(\frac{\left|\mathbf{p}_{i}^{\prime}\right|^{2}}{2m_{i}}-\frac{GMm_{i}}{\left|\mathbf{r}_{i}\right|}\right)\label{eq:ap3}
\end{equation}
that represents the two body motion of the terrestrial bodies around
the Sun, of mass $M$,
\begin{equation}
H_{\mathrm{Pert}}(\mathbf{r})=-\sum_{i=1}^{T-1}\sum_{k=i+1}^{N}\frac{Gm_{i}m_{k}}{\left|\mathbf{r}_{i}-\mathbf{r}_{k}\right|}\label{eq:ap4}
\end{equation}
which is the mutual gravitational perturbation among the terrestrial
bodies,
\begin{equation}
H_{\mathrm{Jov}}(\mathbf{r},t)=-\sum_{i=1}^{N}\sum_{j=1}^{J}\frac{Gm_{i}\mu_{j}}{\left|\mathbf{r}_{i}-\boldsymbol{\rho}_{j}(t)\right|}\label{eq:ap5}
\end{equation}
that gives the direct gravitational perturbation of the giant planets
on the terrestrial bodies, and
\begin{equation}
H_{\mathrm{Sun}}(\mathbf{p}^{\prime},t)=\frac{1}{2M}\left|\sum_{i=1}^{N}\mathbf{p}_{i}^{\prime}+\sum_{j=1}^{J}\boldsymbol{\pi}_{j}^{\prime}(t)\right|^{2}\label{eq:ap6}
\end{equation}
that represents the linear momentum of the Sun around the centre-of-mass
of the whole system.

The detailed sequence of operations for a single time step integration,
from $t$ to $t+\tau$, is as follows:
\begin{description}
\item Step 1. Start with the heliocentric positions and velocities $\mathbf{r},\mathbf{v}$
of all the terrestrial bodies, obtained from the previous time step,
as well as the heliocentric positions and velocities $\boldsymbol{\rho},\boldsymbol{\upsilon}$
interpolated for the giant planets, at time $t$.
\item Step 2. Calculate the heliocentric acceleration $\mathbf{a}_{\mathrm{R}}$
on the terrestrial protoplanets due to relativistic corrections \citep{Quinn1991}:
\begin{equation}
\mathbf{a}_{\mathrm{R},i}  =\frac{GM}{c^{2}\left|\mathbf{r}_{i}\right|^{3}}\left(4\,\mathbf{r}_{i}\cdot\mathbf{v}_{i}\,\mathbf{v}_{i}-\mathbf{v}_{i}\cdot\mathbf{v}_{i}\,\mathbf{r}_{i} +4M\frac{\mathbf{r}_{i}}{\left|\mathbf{r}_{i}\right|}\right),\quad i=1,\ldots,T\label{eq:ap7}
\end{equation}
\item Step 3. Apply a kick, over $\tau/2$, in the heliocentric velocities $\mathbf{v}$
of the terrestrials protoplanets due to the relativistic acceleration
correction:
\begin{equation}
\mathbf{v}_{i}\longleftarrow\mathbf{v}_{i}+\frac{\tau}{2}\mathbf{a}_{\mathrm{R},i},\qquad i=1,\ldots,T\label{eq:ap8}
\end{equation}
\item Step 4. Convert the heliocentric velocities $\mathbf{v},\boldsymbol{\upsilon}$
of all bodies to barycentric $\mathbf{v}^{\prime},\boldsymbol{\upsilon}^{\prime}$,
with respect to the centre-of-mass of the whole system:
\begin{align}
\mathbf{V}^{\prime} & =-{\displaystyle \frac{\sum_{i=1}^{N}m_{i}\mathbf{v}_{i}+\sum_{j=1}^{J}\mu_{j}\boldsymbol{\upsilon}_{j}}{M+\sum_{i=1}^{N}m_{i}+\sum_{j=1}^{J}\mu_{j}}}\nonumber \\
\mathbf{v}_{i}^{\prime} & =\mathbf{v}_{i}+\mathbf{V}^{\prime},\qquad\qquad\qquad i=1,\ldots,N\nonumber \\
\boldsymbol{\upsilon}_{j}^{\prime} & =\boldsymbol{\upsilon}_{j}+\mathbf{V}^{\prime},\,\,\,\quad\qquad\qquad j=1,\ldots,J\label{eq:ap9}
\end{align}
\item Step 5. Compute the heliocentric accelerations $\mathbf{a}_{\mathrm{J}}$
on the terrestrial bodies due to the giant planets:
\begin{equation}
\mathbf{a}_{\mathrm{J},i}=-\sum_{j=1}^{J}G\mu_{j}\frac{\mathbf{r}_{i}-\boldsymbol{\rho}_{j}}{\left|\mathbf{r}_{i}-\boldsymbol{\rho}_{j}\right|^{3}},\qquad i=1,\ldots,N\label{eq:ap10}
\end{equation}
\item Step 6. Apply a kick, over $\tau/2$, in the barycentric velocities $\mathbf{v}^{\prime}$
of the terrestrial bodies due to these accelerations:
\begin{equation}
\mathbf{v}_{i}^{\prime}\longleftarrow\mathbf{v}_{i}^{\prime}+\frac{\tau}{2}\mathbf{a}_{\mathrm{J},i},\qquad i=1,\ldots,N\label{eq:ap11}
\end{equation}
\item Step 7. Perform a linear drift, over $\tau/2$, of the heliocentric positions
$\mathbf{r}$ of the terrestrial bodies due to the giant planets contribution
to the Sun momentum:
\begin{equation}
\mathbf{r}_{i}\longleftarrow\mathbf{r}_{i}+\frac{\tau}{2M}\sum_{j=1}^{J}\mu_{j}\boldsymbol{\upsilon}_{j}^{\prime},\qquad i=1,\ldots,N\label{eq:ap12}
\end{equation}
\item Step 8. Perform a linear drift, over $\tau/2$, of the heliocentric positions
$\mathbf{r}$ of the terrestrial bodies due to their own contribution
to the Sun momentum:
\begin{equation}
\mathbf{r}_{i}\longleftarrow\mathbf{r}_{i}+\frac{\tau}{2M}\sum_{k=1}^{N}m_{k}\mathbf{v}_{k}^{\prime},\qquad i=1,\ldots,N\label{eq:ap13}
\end{equation}
\item Step 9. Compute the heliocentric mutual accelerations $\mathbf{a}_{\mathrm{T}}$
of the terrestrial protoplanets and planetesimals:
\begin{align}
\mathbf{a}_{\mathrm{T},i} & ={\displaystyle -\sum_{k=i+1}^{N}Gm_{k}\frac{\mathbf{r}_{i}-\mathbf{r}_{k}}{\left|\mathbf{r}_{i}-\mathbf{r}_{k}\right|^{3}}},\qquad i=1,\ldots,T\nonumber \\
\mathbf{a}_{\mathrm{T},k} & ={\displaystyle -\sum_{i=1}^{T}Gm_{i}\frac{\mathbf{r}_{k}-\mathbf{r}_{i}}{\left|\mathbf{r}_{k}-\mathbf{r}_{i}\right|^{3}}},\qquad k=T+1,\ldots,N\label{eq:ap14}
\end{align}
\item Step 10. Apply a kick, over $\tau/2$, in the barycentric velocities $\mathbf{v}^{\prime}$
of the terrestrial bodies due to these accelerations:
\begin{equation}
\mathbf{v}_{i}^{\prime}\longleftarrow\mathbf{v}_{i}^{\prime}+\frac{\tau}{2}\mathbf{a}_{\mathrm{T},i},\qquad i=1,\ldots,N\label{eq:ap15}
\end{equation}
\item Step 11. Drift the heliocentric positions $\mathbf{r}$ and barycentric velocities
$\mathbf{v}^{\prime}$ of the terrestrial protoplanets and planetesimals
along the Keplerian orbits generated by the Hamiltonian $H_{\mathrm{Kep}}$
(Eq. \ref{eq:ap3}), over a full time step $\tau$
\item Step 12. Recompute the heliocentric mutual accelerations $\mathbf{a}_{\mathrm{T}}$
of the terrestrial bodies (Eq. \ref{eq:ap14})
\item Step 13. Apply a kick, over $\tau/2$, in the barycentric velocities $\mathbf{v}^{\prime}$
of the terrestrial bodies due to these accelerations (Eq. \ref{eq:ap15})
\item Step 14. Perform a linear drift, over $\tau/2$, of the heliocentric positions
$\mathbf{r}$ of the terrestrial bodies due to their own contribution
to the Sun momentum (Eq. \ref{eq:ap13})
\item Step 15. Convert the heliocentric velocities $\boldsymbol{\upsilon}$ of the
giant planets at time $t+\tau$ to barycentric $\boldsymbol{\upsilon}^{\prime}$,
with respect to the centre-of-mass of the whole system:
\begin{align}
\mathbf{V}^{\prime} & ={\displaystyle -\frac{\sum_{i=1}^{N}m_{i}\mathbf{v}_{i}^{\prime}+\sum_{j=1}^{J}\mu_{j}\boldsymbol{\upsilon}_{j}}{M+\sum_{j=1}^{J}\mu_{j}}}\nonumber \\
\boldsymbol{\upsilon}_{j}^{\prime} & =\boldsymbol{\upsilon}_{j}+\mathbf{V}^{\prime},\qquad\qquad\qquad j=1,\ldots,J\label{eq:ap16}
\end{align}
\item Step 16. Perform a linear drift, over $\tau/2$, of the heliocentric positions
$\mathbf{r}$ of the terrestrial bodies due to the giant planets contribution
to the Sun momentum (Eq. \ref{eq:ap12})
\item Step 17. Recompute the heliocentric accelerations $\mathbf{a}_{\mathrm{J}}$
on the terrestrial bodies due to the giant planets (Eq. \ref{eq:ap10})
\item Step 18. Apply a kick, over $\tau/2$, in the barycentric velocities $\mathbf{v}^{\prime}$
of the terrestrial bodies due to these accelerations (Eq. \ref{eq:ap11})
\item Step 19. Convert back the barycentric velocities $\mathbf{v}^{\prime}$ of
the terrestrial bodies to heliocentric $\mathbf{v}$:
\begin{align}
\mathbf{V}^{\prime} & ={\displaystyle -\frac{\sum_{i=1}^{N}m_{i}\mathbf{v}_{i}^{\prime}+\sum_{j=1}^{J}\mu_{j}\boldsymbol{\upsilon}_{j}^{\prime}}{M}}\nonumber \\
\mathbf{v}_{i} & =\mathbf{v}_{i}^{\prime}-\mathbf{V}^{\prime},\qquad\qquad\qquad i=1,\ldots,N\label{ap17}
\end{align}
\item Step 20. Recalculate the heliocentric acceleration $\mathbf{a}_{\mathrm{R}}$
on the terrestrial protoplanets due to relativistic corrections (Eq.
\ref{eq:ap7})
\item Step 21. Apply a kick, over $\tau/2$, in the heliocentric velocities $\mathbf{v}$
of the terrestrials protoplanets due to the relativistic acceleration
correction (Eq. \ref{eq:ap8})
\item Step 22. Return to step 2.
\end{description}
At variance with the standard \texttt{SyMBA} code, that uses a LD - K - D - K - LD
sequence, \texttt{iSyMBA} uses a K$_{\mathrm{J}}$ - LD$_{\mathrm{J}}$ - LD$_{\mathrm{T}}$ - K$_{\mathrm{T}}$ - D$_{\mathrm{T}}$ - K$_{\mathrm{T}}$ - LD$_{\mathrm{T}}$ - LD$_{\mathrm{J}}$ - K$_{\mathrm{J}}$
sequence, where the sub-indices J,T refer to jovian and terrestrial,
respectively, with an additional outer K$_{\mathrm{R}}$ sequence
for relativistic corrections, if required. This specific sequence
has three advantages over other possible sequences:
\begin{enumerate}
\item The giant planets perturbations (Eq. \ref{eq:ap10}) do not
need to be recomputed at the beginning of each time step; they can
be recovered from the final values of the previous time step.
\item If relativistic corrections are not accounted for, there is
no need to recompute the barycentric velocities at the beginning of
each time step (Eq. \ref{eq:ap9}); they can be recovered from the
final values of the previous time step.
\item The innermost LD$_{\mathrm{T}}$ - K$_{\mathrm{T}}$ - D$_{\mathrm{T}}$ - K$_{\mathrm{T}}$ - LD$_{\mathrm{T}}$
sequence (steps 8 to 14), that treats the interactions among the terrestrial
bodies solely, could be executed by the standard \texttt{SyMBA} algorithm.
\end{enumerate}

\subsubsection{Treatment of close encounters among terrestrial bodies}\label{closeenc}

Close encounters among terrestrial protoplanets, or between terrestrial
protoplanets and planetesimals, are manipulated by \texttt{iSyMBA}
using the same strategy of \texttt{SyMBA}, i.e. by splitting the potential
term (Eq. \ref{eq:ap4}) as:
\begin{align}
H_{\mathrm{Pert}}(\mathbf{r}) & =H_{\mathrm{Pert}}^{(0)}+\sum_{s=1}^{n}H_{\mathrm{Pert}}^{(s)}\\
H_{\mathrm{Pert}}^{(0)} & =-\sum_{i=1}^{T-1}\sum_{k=i+1}^{N}\frac{Gm_{i}m_{k}}{r_{ik}}\prod_{l=1}^{n}\left(1-w_{l,ik}\right)\\
H_{\mathrm{Pert}}^{(s)} & =-\sum_{i=1}^{T-1}\sum_{k=i+1}^{N}\frac{Gm_{i}m_{k}}{r_{ik}}w_{s,ik}\prod_{l=1}^{s-1}\left(1-w_{l,ik}\right)
\end{align}
where the $w_{l,ik}$ are given by Eq. (\ref{eq:ap19-2}). The propagation
of the orbits of any two bodies involved in an encounter during the
innermost LD$_{\mathrm{T}}$-K$_{\mathrm{T}}$-D$_{\mathrm{T}}$-K$_{\mathrm{T}}$-LD$_{\mathrm{T}}$
sequence (steps 8 to 14), is carried out recursively through the nested
application of a K$_{\mathrm{T}}$-D$_{\mathrm{T}}$-K$_{\mathrm{T}}$
sequence to the Hamiltonian $\left(\left(\left(H_{\mathrm{Kep}}+H_{\mathrm{Pert}}^{(0)}\right)+H_{\mathrm{Pert}}^{(1)}\right)+\ldots\right)+H_{\mathrm{Pert}}^{(n)}$,
using smaller and smaller time steps $\tau_{l}=\tau/3^{l}$, $l=1,\ldots,n$.

For this stage, we use the original \texttt{SyMBA} subroutines, with
little modifications. Bodies are always merged whenever they
reach the last stage of the recursion and get closer than $R_{n+1,ik}$.

\subsubsection{Treatment of close encounters between terrestrial bodies and giant
planets}

Close encounters with giant planets are not properly manipulated by
\texttt{iSyMBA}. The application of a splitting strategy to the term
$H_{\mathrm{Jov}}$ (Eq. \ref{eq:ap5}) would demand additional interpolations
of the giant planets orbits, down to smaller and smaller time steps.
This would also require significant changes to the standard \texttt{SyMBA}
recursive subroutines, which turns to be a quite complex task. Therefore,
we leave this implementation to a future work

Currently, the way \texttt{iSyMBA} deals with such close encounters
is to keep tracking the distances between the protoplanets/planetesimals
and the giant planets, and discard the former when they get closer
to a giant planet than the sum of their individual Hill radii. This
solution proved to be adequate for our purposes, since the disk of
terrestrial bodies is interior to Jupiter's orbit. Thus, we may expect
that close encounters of planetesimals with the giant planets will
be rare, and they eventually will affect only the outer edge of the
disk. In particular, using the results of \citet{Nesvorny2021}, we verified that, when the disk extends up to 4 au
with a radial surface density profile $\Sigma(r)=r^{-1}$, less than
30\% of the planetesimals experienced close encounters with Jupiter
during the simulations.

\subsubsection{Treatment of small perihelion passages}

The \texttt{iSyMBA} code does not properly treat the $H_{\mathrm{Sun}}$ term increase (Eq. \ref{eq:ap6}) when a terrestrial
body experiences a small perihelion passage or get too close to the
Sun. This limitation also exists in the standard \texttt{SyMBA} code.
In such cases, the code relies only in the setup of a sufficiently
small time step to resolve the perihelion passage without loosing
too much precision. In \citet{Nesvorny2021}, we setup a time step $\tau=3$-7
days, which proved to be good enough in that application. Bodies are discarded whenever
they reach heliocentric or perihelion distances smaller than a given threshold, set by the user.

\subsubsection{Symplecticity and energy conservation}

The symplecticity of \texttt{iSyMBA} is guaranteed by the conservation of the corresponding surrogate Hamiltonian:
\begin{align}
\overline{H} & =H_{\mathrm{Kep}}+H_{\mathrm{Pert}}+H_{\mathrm{Jov}}+H_{\mathrm{Sun}} \nonumber \\
& -\sum_{i=0}^{n}\frac{\tau_{i}^{2}}{12} \Big[\Big[ H_{\mathrm{Kep}}\,,\, H_{\mathrm{Pert},i} \Big]\,,\, H_{\mathrm{Kep}} +\frac{1}{2}H_{\mathrm{Pert},i}+\sum_{j=i+1}^{n}H_{\mathrm{Pert},j} \Big] \nonumber \\
& -\frac{\tau_{0}^{2}}{12} \Big[\Big[ H_{\mathrm{Kep}}\,,\, H_{\mathrm{Sun}} \Big]\,,\, H_{\mathrm{Kep}}+\frac{1}{2}H_{\mathrm{Sun}} \Big] \nonumber \\
& -\frac{\tau_{0}^{2}}{12} \Big[\Big[ H_{\mathrm{Kep}}\,,\, H_{\mathrm{Jov}} \Big]\,,\, H_{\mathrm{Kep}}+\frac{1}{2}H_{\mathrm{Jov}} \Big] +\mathcal{O}(\tau^{4}) \label{eq:Hsurr}
\end{align}
where $[,]$ represents the Lagrange brackets. In practice, the total energy can be computed as the sum of the barycentric kinetic and potential energies, taking
into account all the three categories of bodies in the system:
\begin{align}
E & =\frac{1}{2}M\big|\mathbf{V}^{\prime}\big|^{2}+\sum_{i=1}^{N}\frac{1}{2}m_{i}\big|\mathbf{v}_{i}^{\prime}\big|^{2}+\sum_{j=1}^{J}\frac{1}{2}\mu_{j}\big|\boldsymbol{\upsilon}_{j}^{\prime}\big|^{2}\nonumber \\
 & -\sum_{i=1}^{N}\frac{GMm_{i}}{\big|\mathbf{r}_{i}\big|}-\sum_{i=1}^{T-1}\sum_{k=i+1}^{N}\frac{Gm_{i}m_{k}}{\big|\mathbf{r}_{i}-\mathbf{r}_{k}\big|}-\sum_{j=1}^{J}\frac{GM\mu_{j}}{\big|\boldsymbol{\rho}_{j}\big|}\nonumber \\
 & -\sum_{j=1}^{J-1}\sum_{k=j+1}^{J}\frac{G\mu_{j}\mu_{k}}{\big|\boldsymbol{\rho}_{j}-\boldsymbol{\rho}_{k}\big|}-\sum_{i=1}^{N}\sum_{j=1}^{J}\frac{Gm_{i}\mu_{j}}{\big|\mathbf{r}_{i}-\boldsymbol{\rho}_{j}\big|}\label{eq:ene}
\end{align}
It is worth recalling that in \texttt{iSyMBA}, the model Hamiltonian
is non autonomous. Therefore, the energy is not expected to be constant, but it should display periodic oscillations with bounded amplitude.

\subsection{Validation tests}

The \texttt{iSyMBA} algorithm has been validated as follows. We consider a system consisting of $J$ giant planets and $T$ terrestrial planets, and perform a simulation over 1 My using the standard \texttt{SyMBA}, i.e. allowing
for all the planets to be mutually perturbed. The output (orbital elements) of this simulation
for the giant planets is stored in a file every 1 year, while the output for the terrestrial planets is stored in a separate file every 1\,000 years.
Then, we repeat the simulation using \texttt{iSyMBA}, with the same initial conditions for the terrestrial planets, and interpolating
the giant planets' orbits from the ones previously stored. The output of this simulation is recorded every 1\,000 years, and it is directly compared to the previous output from \texttt{SyMBA}. In all the simulations, we check that the total energy of the system, computed from Eq. (\ref{eq:ene}), is well behaved. 

We apply the above validation test to two different systems. The first one is the present solar system, with 4 terrestrial and 4 giant planets. Initial conditions are taken from the JPL Ephemerides. Relativistic perturbations are not taken into account, and there are no close encounters between the planets. We call this the 4G4T model.

The second system is a fictitious system composed of 5 giant planets and 20 terrestrial bodies. The giant planets are Jupiter, Saturn, and 3 ice giants, initially in a mutual resonant and compact orbital configuration \citep[see][for example]{DeSouza2021}. The 20 terrestrial bodies are represented by planetary embryos with a total mass of 5~$M_\oplus$. These bodies are uniformly distributed in a very narrow annulus, between 0.95 and 1.05 au, with eccentricities $<0.01$ and inclinations $<0.001$. The idea of this setup is to force close encounters between the terrestrial bodies, in order to test the behaviour of \texttt{iSyMBA} under such conditions. Relativistic perturbations are not taken into account either. We call this the 5G20T model.

\begin{figure*}
\centering
\includegraphics[width=\textwidth]{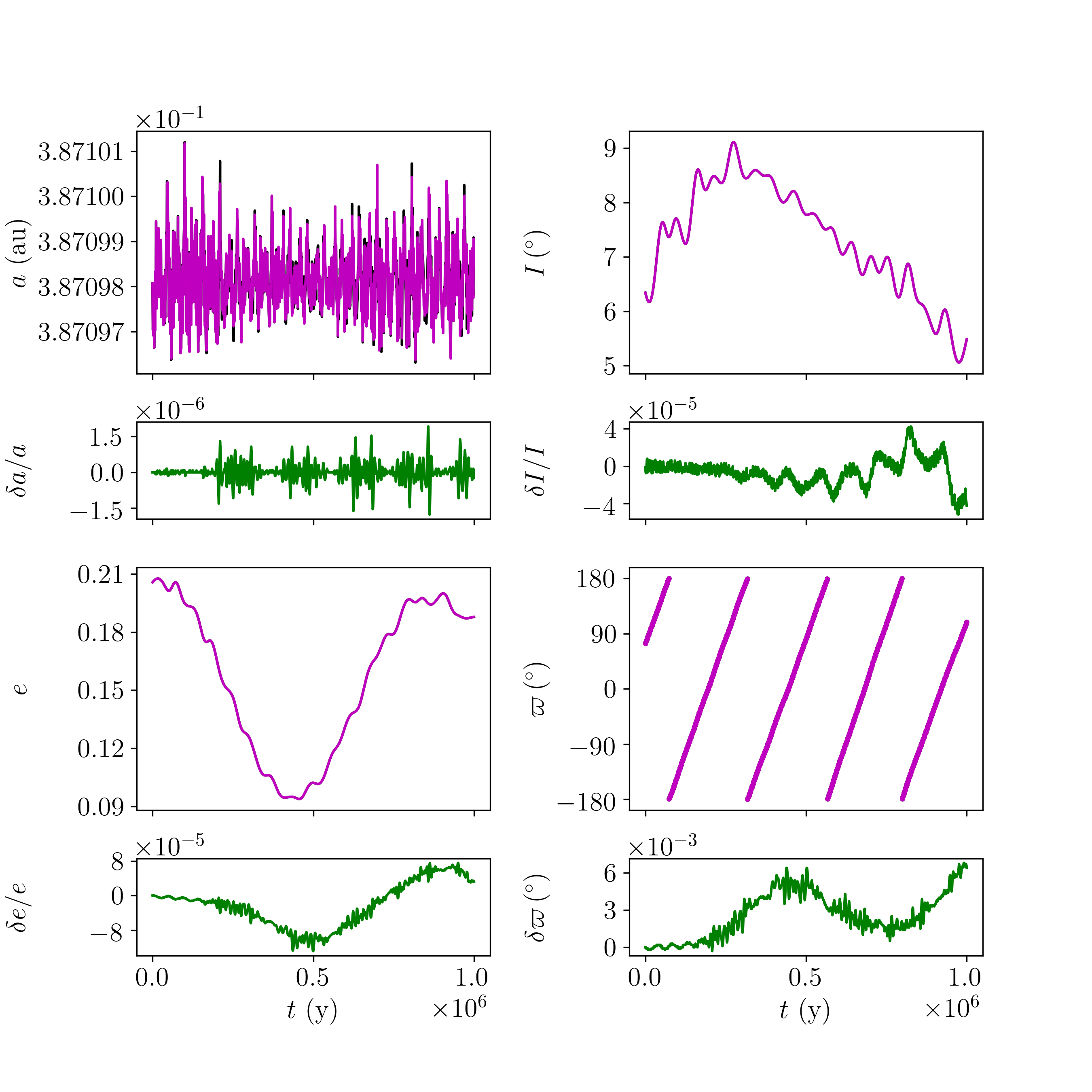}
\caption{Comparison between the output from \texttt{SyMBA} (black) and \texttt{iSyMBA} (over-plotted in magenta) for the orbital elements of planet Mercury, over 1 My of evolution, in the 4G4T model (see text for the simulation details): $a$, semi-major axis, $e$ eccentricity, $I$ inclination, and $\varpi$ longitude of perihelion. The smaller panels show the \textbf{relative} differences (in green), except for $\varpi$ that shows the absolute difference.}
\label{fig3}
\end{figure*}

Figure \ref{fig3} shows a result from the 4G4T model. The panels display the evolution of the orbital elements of Mercury from the iSyMBA simulation (in magenta), and the SyMBA simulation (in black, but not visible due to overlapping). The differences between the two codes are shown in green. The behaviour is similar for the other terrestrial planets. Table \ref{tab:relative} summarises the maximum relative differences found in this validation test. The total energy of the system behaves as expected, and the differences between the two codes in no larger than $10^{-5}$ over the whole time span.

\begin{table}
    \centering
    \caption{Maximum differences in the orbital elements of the terrestrial planets, over 1 My, for the validation test using the 4G4T model (see text). }
    \label{tab:relative}
    \begin{tabular}{ccccc}
    \hline
         & $\delta a/a$ & $\delta e/e$ & $\delta I/I$ & $\delta\varpi\,(^{\circ})$ \\
    \hline
        Mercury & $1.5\times 10^{-6}$ & $9\times 10^{-5}$ & $4\times 10^{-5}$ & $6\times 10^{-3}$ \\
        Venus & $2.5\times 10^{-5}$ & $6\times 10^{-3}$ & $2\times 10^{-3}$ & $2\times 10^{-1}$ \\
        Earth & $4\times 10^{-5}$ & $1.5\times 10^{-2}$ & $6\times 10^{-3}$ & $5\times 10^{-1}$ \\
        Mars & $3\times 10^{-5}$ & $1.5\times 10^{-3}$ & $2\times 10^{-4}$ & $4\times 10^{-2}$ \\
    \hline
    \end{tabular}
\end{table}

In the 5G20T model, reproducing the exact evolution of the system with \texttt{iSyMBA} is not feasible, because the system is chaotic, and the small differences caused by the many collisions/mergers produce quantitatively different results. We verify, however, that the total energy of the system is well behaved, as shown in Fig. \ref{fig5}, with a bounded amplitude for \texttt{iSyMBA} that is only twice than in the 4G4T model without collisions. Although the number of collisions recorded in \texttt{SyMBA} and \texttt{iSyMBA} is not the same, the latter is able to reproduce quite well the first few collisions in the simulation. An example of this is shown in Fig. \ref{fig6}. We note that the merger happens at slightly different times in each case. Such small differences quickly propagate, and in a few tens of years each code starts to produce its own set of collisions, not matching each other anymore.

\begin{figure}
\centering
\includegraphics[width=\columnwidth]{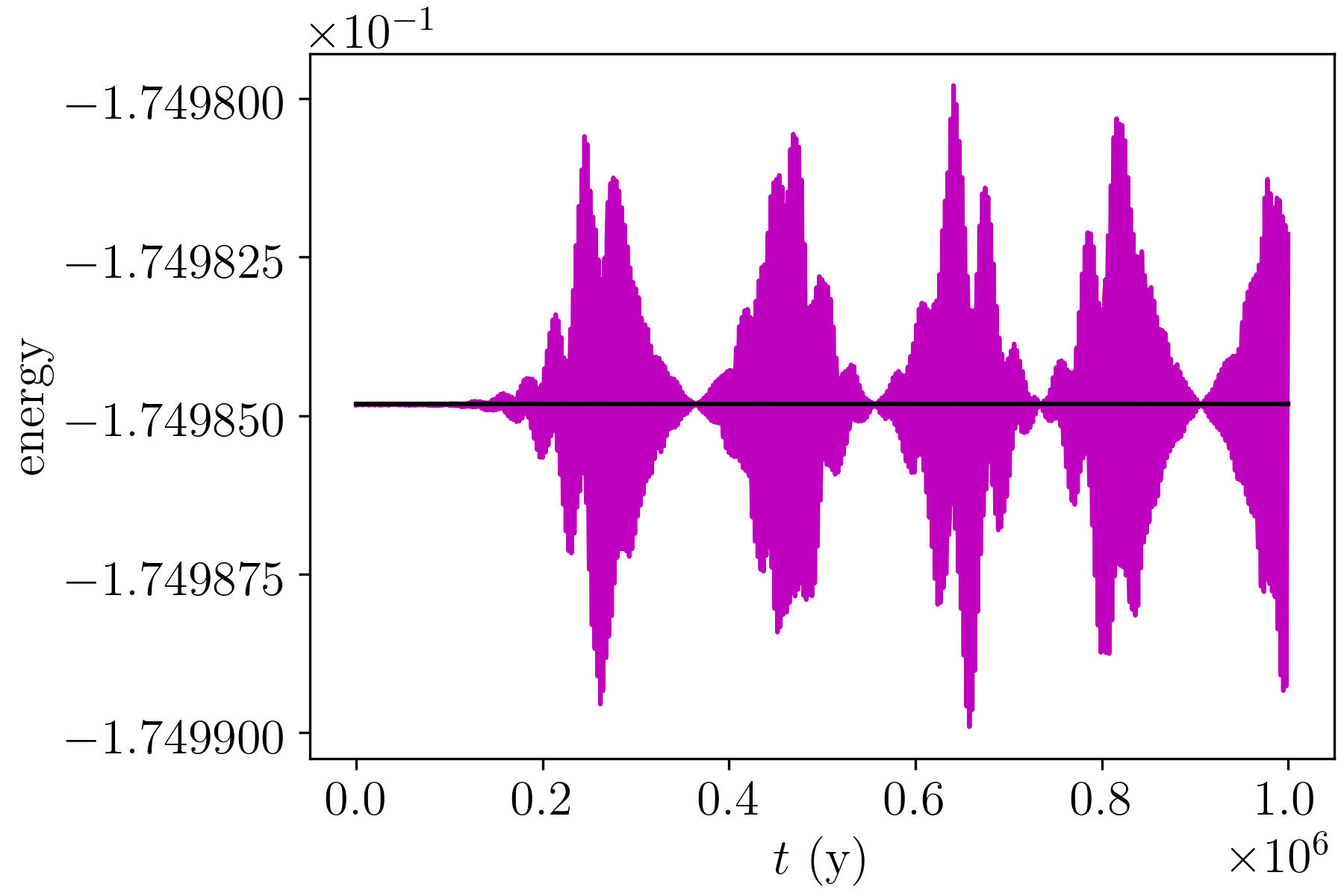}
\caption{Comparison between the total energy of the system from \texttt{SyMBA} (black) and from \texttt{iSyMBA} (magenta), in the 4G4T model (see text). Since the \texttt{iSyMBA} Hamiltonian is non autonomous, the energy displays oscillations with amplitude $<10^{-5}$, with respect to the \texttt{SyMBA} energy.}
\label{fig4}
\end{figure}

\begin{figure}
\centering
\includegraphics[width=\columnwidth]{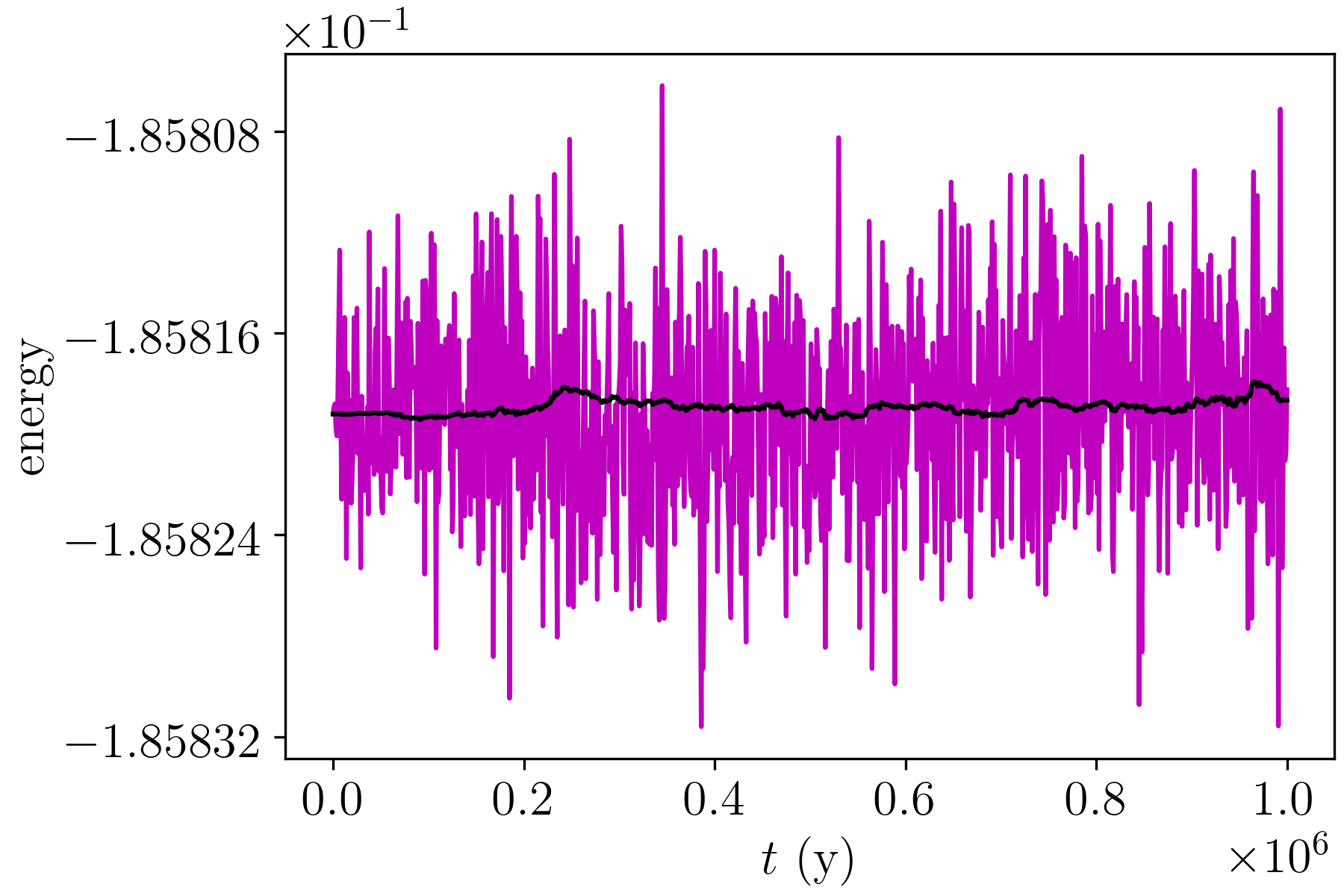}
\caption{Comparison between the total energy of the system from \texttt{SyMBA} (black) and from \texttt{iSyMBA} (magenta), in the 5G20T model (see text). 
Again, the \texttt{iSyMBA} energy displays oscillations with amplitude $<2\times 10^{-5}$, with respect to the \texttt{SyMBA} energy.}
\label{fig5}
\end{figure}

\begin{figure}
\centering
\includegraphics[width=\columnwidth]{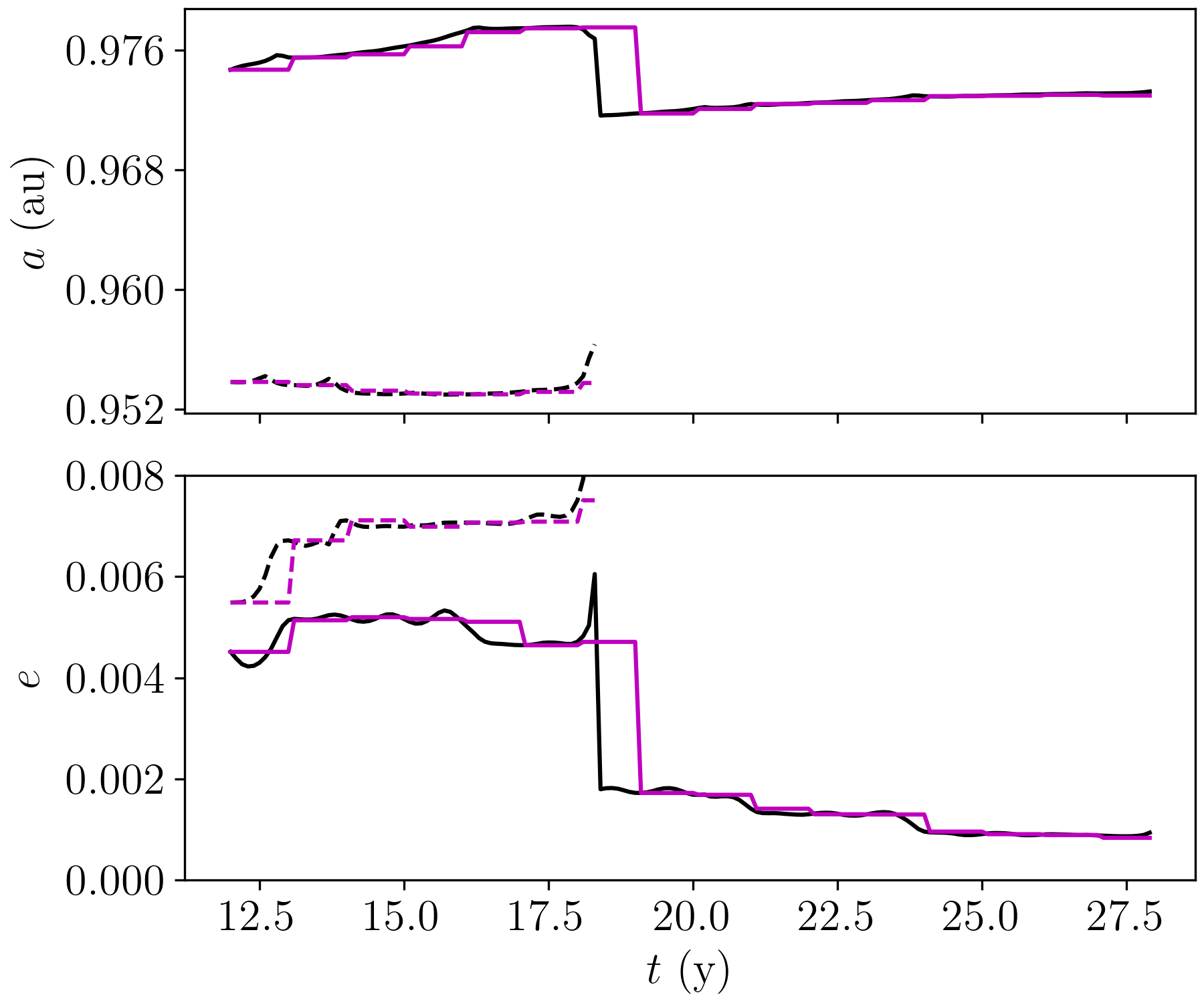}
\caption{Detail of an encounter/merger between two terrestrial bodies in the 5G20T model. The \texttt{SyMBA} simulation is shown in black, and \texttt{iSyMBA} in magenta. Full and dashed lines identify each of the bodies, respectively.}
\label{fig6}
\end{figure}

The above validation tests allow us to conclude that \texttt{iSyMBA} shows the desired behaviour in terms of simulation results.

\section{Parallelisation strategy}\label{setc3}

\begin{figure}
\centering
\includegraphics[width=\columnwidth]{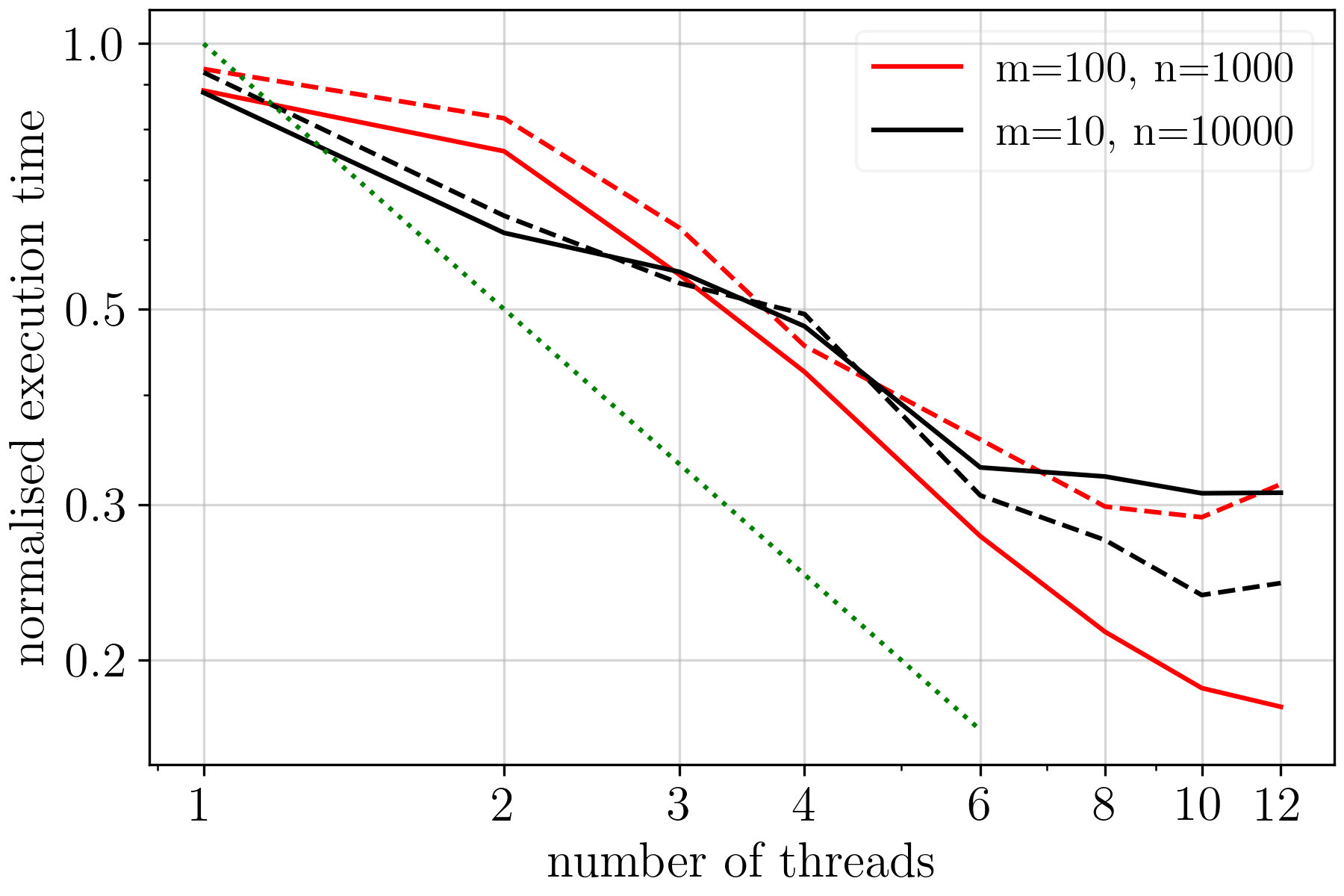}
\caption{The performance of \texttt{iSyMBA} for different multi-threading parallelisation strategies. The solid lines correspond to the parallelisation of the outer loop in double loops, while the dashed lines correspond to the parallelisation of the inner loop. The colours identify the test models with different numbers of proto-planets (m) and planetesimals (n). The execution time has been normalised with respect to the serial execution time (see text for details). The green dotted line represents a linear trend, for reference purposes.}
\label{fig7}
\end{figure}

Aiming to improve the performance of \texttt{iSyMBA}, we have implemented multi-threading parallelisation using \texttt{OpenMP}. The following discussion does not intend to be comprehensive, and it is only applicable to the specific test models described below. Our aim is to provide some clues about the possible best strategies to parallelise our code, that eventually can be also taken into account to parallelise \texttt{SyMBA} itself. 

There is no unique strategy to parallelise a code, and sometimes the best approach is obtained by first redesigning the original serial code. Here, however, the idea is to adopt a parallelisation strategy that keeps the original serial structure of the \texttt{SyMBA} subroutines with minimum or no changes.

The following discussion is based on two different test models: one considering 100 terrestrial proto-planets and 1\,000 planetesimals \citep[e.g.][]{Nesvorny2021}, which we refer to as m100n1000 model, and another considering 10 proto-planets and 10\,000 planetesimals, which we called m10n10000 model. In both cases the number of giant planets is 5. In all the tests, we use the same total time span and the same time step. We also keep the amount of I/O operations to the minimum required. We define the normalised execution time as the ratio $T_\mathrm{par}/T_\mathrm{ser}$, where $T_\mathrm{ser}$ is the total execution time of the serial code, without any parallelisation, and $T_\mathrm{par}$ is the execution time of the parallelised code. All the tests have been performed in Intel Core i7 processors, using \texttt{GNU Fortran}.  

There are basically two types of structures that can be parallelised in \texttt{iSyMBA} using \texttt{OpenMP}:
\begin{enumerate}
    \item the operations that require double loops over the terrestrial bodies, like the mutual acceleration calculations (Eqs. \ref{eq:ap7}, \ref{eq:ap10} and \ref{eq:ap14}), the check for close encounters, and the energy computation (Eq. \ref{eq:ene}), and 
    \item the calculations that require single loops over the bodies, including the Keplerian drifts, the linear drifts, the kicks, the coordinate changes, the loops to deal with interpolation of the giant planets, etc.
\end{enumerate}
We will discuss each structure separately.

\subsection{Single loops}

Paralellisation of the single loops within the code has to be carefully evaluated, because it may provide little or no improvement of the execution speed. For example, in the LD - K - D - K - LD integration scheme, the Keplerian drifts are, in theory, the second most CPU-expensive step, after the accelerations calculation. However, in our simulations, the Keplerian drifts take between 5\% to 20\% of the total run time in a serial run. Therefore, their parallelisation might not contribute significantly to improve performance.

We have verified that, in the m100n1000 simulations, parallelisation of the single loops makes the code only $\sim 1.2$ times faster, but in the m10n10000 simulations, it becomes $\sim 2.3$ times faster. 

We have also verified that parallelisation of any loop within the recursive integration of close encounters (see Sect. \ref{closeenc}) must be avoided, since it may slow down execution speed by a factor of 2. This concerns, in particular, the subroutines \texttt{symba7\_step\_recur} and \texttt{symba7\_kick}.  
We note, however, that there is no actual need to parallelise any part of the recursion, because it only affects the pair of bodies involved in an encounter, and these are not too frequent per time step. 

After several experiments, we conclude that multi-threading parallelisation has to focus on the double loops, as explained below, and on the single loops that performs the most complex calculations, like the Keplerian drifts, the loops to deal with the interpolation of the giant planets, the relativistic corrections, the oblateness potential, and the discard subroutines.

\subsection{Double loops}

The strategy applied for parallelisation of the double loops influences the execution speed. A double loop to compute the accelerations between proto-planets and planetesimals typically reads:

\texttt{do i from 1 to m}

\texttt{\quad{}do j from i+1 to n}

\texttt{\qquad{}a(j) = a(j) + accel\_ji}

\texttt{\qquad{}a(i) = a(i) + accel\_ij}

\texttt{\quad{}end do}

\texttt{end do}

\noindent where \texttt{m} is the number of self gravitating proto-planets, \texttt{n} is the number of planetesimals, and \texttt{a()} is an array of dimension \texttt{n}. In this case, one possible strategy is to parallelise the outer loop over \texttt{i}, and the other possibility is to parallelise the inner loop over j.

Figure \ref{fig7} shows the normalised execution time for the different models and strategies, as a function of the number of threads. The solid lines correspond to the case in which the outer loops are parallelised, while the dashed lines correspond to the case in which the inner loops are parallelised.  
We note that, when using few threads, there is no significant difference between one strategy or the other, although parallelisation of the outer loops performs a bit better. On the other hand, when using many threads, each strategy has advantages over the other depending on the values of \texttt{m} and \texttt{n}.

For the simulations of the m100n1000 model, parallelisation of the outer loops provides $\sim 1.5$ times faster execution times with respect to the parallelisation of the inner loops. On the other hand, for the simulations of the m10n10000 model, it is the parallelisation of the inner loops that provides $\sim 1.3$ times faster execution times than parallelising the outer loops.\footnote{These tests have been performed using 8 threads.} 

This behaviour may be related to the fact that initialisation and execution of a parallel loop involve several tasks, besides the calculations within the loop, that may produce some latency in the execution. When parallelising an inner loop, the parallel threads have to be initialised/allocated for every iteration of the outer loop, and this may cause a lot of latency. If the outer loop is small, as in \,\texttt{m} = 10, the latency does not have a big impact. However, if the outer loop is bigger, as in \,\texttt{m} = 100, the impact of latency may become significant.

Parallelisation of the outer loops has a couple of additional peculiarities that should be taken into account. The first one refers to the fact that the number of iterations over \texttt{j}, in the inner loop, is smaller for larger \texttt{i}. This means that the amount of work to be done by each iteration over \texttt{i} is different. In such case, the way of scheduling the iterations may be relevant. 
One possibility is choosing between a cyclic or a block scheduling. A cyclic schedule distributes the loop iterations in a round-robin fashion among the available threads, and should provide better performance in our case. The other possibility is choosing between static or dynamic scheduling. We have verified that using a combination of static and cyclic scheduling provides $\sim 1.1$ times faster execution times than either a dynamic or block scheduling.

The second peculiarity in parallelising the outer loops refers to the occurrence of data racing conditions over the accelerations array \texttt{a()}. A racing condition arises when two or more processes running in different threads try to modify or update the same variable at the same time. Fortunately, \texttt{Fortran OpenMP} has the capability of performing array reduction, which allows to properly update \texttt{a()}, avoiding data race.

Although multi-threading parallelisation, in our case, may improve execution speed by a factor of 3 to 6, the improvement is not linear with the number of threads and, as shown in Fig. \ref{fig7}, it tends to stabilise for $\gtrsim 10$ threads.  

\section{Conclusions}\label{sect4}

In this work, we described how to implement the necessary
modifications to embed an orbit interpolation scheme into the symplectic
planetary $N$-body integrator \texttt{SyMBA}. Our algorithm, named
\texttt{iSyMBA}, allows to study the effects of a prescribed evolution
of a set of planets on a target population of massive bodies, that
interact with each other through close encounters. 

\texttt{iSyMBA} is a very useful code to accurately evaluate the effects of planetary instabilities on the accretion processes in the terrestrial
planets region. These include the growth of protoplanetary embryos
\citep{Nesvorny2021}, the Moon-forming impact \citep{DeSouza2021}, and the origin of
Mercury.

Although \texttt{iSyMBA} has been primarily developed and implemented to study terrestrial planet formation, the method presented here could be easily modified to study the evolution of other populations, that requires orbit interpolation from previously developed simulations, while accounting for close encounters among massive objects.

\section*{Acknowledgements}

We wish to thank an anonymous referee for helpful comments and suggestions. FR acknowledges support form the Brazilian National Council of Research (CNPq). The simulations have been performed at the SDumont cluster of the Brazilian System of High-Performance Computing (SINAPAD)

\section*{Data Availability}

\texttt{iSyMBA} is freely available under request to the corresponding author.


\bibliographystyle{mnras}
\bibliography{isymba}

\bsp	
\label{lastpage}
\end{document}